\documentclass[reprint,amsmath,amssymb,prb,longbibliography]{revtex4-2}

\usepackage{graphicx}
\usepackage[utf8]{inputenc}
\usepackage{xcolor}
\definecolor{goodgreen}{rgb}{0.1,0.5,0}
\definecolor{goodred}{rgb}{0.7,0,0}
\usepackage[colorlinks,urlcolor=goodgreen,citecolor=blue,linkcolor=goodred]%
{hyperref}
\usepackage{upgreek}
\usepackage{newtxtext,newtxmath}

\newcommand{\vg}{\ensuremath{V_\text{G}}}
\newcommand{\cg}{\ensuremath{C_\text{G}}}

\newcommand{\ctot}{\ensuremath{C_{\Sigma}}}

\newcommand{\vsd}{\ensuremath{V_\text{SD}}}

\newcommand{\un}[1]{\ensuremath{\,\text{#1}}}

\begin{document}

\title{Anomalous Duffing mechanics of a suspended carbon nanotube quantum dot at
ultrastrong coupling}

\author{Akong N. Loh}
\email{akong.loh@ur.de}
\author{Furkan R. Özyi\u{g}it}
\author{Fabian Stadler}
\author{Katrin Burkert}
\author{Niklas Hüttner}
\author{Andreas K. Hüttel}
\email{andreas.huettel@ur.de}
\affiliation{Institute for Experimental and Applied Physics, 
University of Regensburg, Universitätsstr.\ 31, 93053 Regensburg, 
Germany}

\date{\today}

\begin{abstract}
At cryogenic temperatures, suspended single-wall carbon nanotube quantum dots 
act both as prototypical quantum dots as well as high-quality factor mechanical 
resonators. Single-electron tunneling enables reaching an ultrastrong 
electron-vibron coupling regime, where the coupling parameter exceeds the 
vibration frequency. Due to the high quality factors, a strongly nonlinear 
Duffing response is easily reached. Here, we quantitatively study the Duffing 
response parameters of such a device and their relation to Coulomb blockade 
oscillation. At the edges of single-electron tunneling regions, a local 
increase of the Duffing parameter corresponding to a stiffening spring is 
observed. Size and approximate scaling of the effect agree with single-electron 
tunneling phenomena, which however should lead to softening spring behaviour. 
Possible causes of these puzzling results are discussed.
\end{abstract}

\maketitle 

% \section{Introduction}

The outstanding mechanical and electronic properties of single-wall carbon 
nanotubes (SW-CNTs) have made them a material of choice for studying 
nanoelectromechanical systems at the macromolecular scale 
\cite{nature-sazonova-2004, highq, nnano-moser-2014, rmp-laird-2015,
rmp-bachtold-2022, arxiv-sevitz-2025}. Their low mass, nanoscale size, extreme 
aspect ratio, and large mechanical quality factors at cryogenic temperatures 
have been exploited for ultrasensitive force, mass, and charge sensing 
\cite{nl-lassagne-2008, nl-hakkinen-2015, nl-debonis-2018}. Furthermore, the 
interaction between single-electron tunneling (SET) in a CNT quantum dot and 
the motion of the CNT has been studied extensively \cite{strongcoupling, 
s-lassagne-2009, prb-meerwaldt-2012, heliumdamping, deng_strongly_2016, 
nphys-urgell-2020, optomechanics, nphys-samanta-2023, arxiv-sevitz-2025}. 
Strong electron-vibron coupling and nonlinear mechanics of CNTs offer exciting 
possibilities for diverse devices, including nano-electromechanical 
qubits \cite{prx-pistolesi-2021}, quantum state transfer between degrees of 
freedom, high-precision sensors, sideband cooling and driving, and in general 
the exploration of fundamental questions in quantum mechanics 
\cite{strongcoupling, optomechanics, nphys-wen-2020, prx-pistolesi-2021, 
nphys-samanta-2023}.

In this work, we study the mechanical response of a SW-CNT nanomechanical 
resonator with an embedded quantum dot. Back action of tunneling electrons
causes a softening of the mechanical mode \cite{strongcoupling, s-lassagne-2009} 
and demonstrates ultrastrong coupling between vibration and single-electron
tunneling \cite{prr-vigneau-2022}. We obtain an electromechanical coupling
parameter $g / 2\pi = 2.3\un{GHz}$ and a coupling ratio $g / \omega_\text{m}=4.0$
putting us deep in the ultrastrong coupling regime. Between the Coulomb 
oscillations of conductance, as a consequence of high mechanical quality factor 
\cite{prb-meerwaldt-2012} and low dynamic range \cite{apl-postma-2005}, a 
nonlinear Duffing response of the electromechanical system 
\cite{book-duffing-1918} can be observed over a wide driving power range. 
Fitting the jump-down points of the response curve with a backbone equation, 
both linear and nonlinear (Duffing parameter) components of the CNT spring 
constant are extracted. The evolution of the nonlinear response is studied for 
a range of gate voltages across a Coulomb oscillation. Surprisingly, while the 
absolute value $\left|\beta(\vg)\right|$ of the Duffing parameter scales as 
expected from single-electron tunneling interactions \cite{strongcoupling}, we 
observe a stiffening spring, that is, its sign differs from expectation. 

% \section{Device characterization}

\begin{figure}
\includegraphics{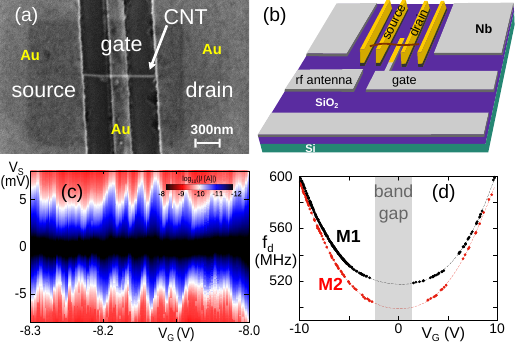}
\caption{\label{fig-device}
(a) Top view scanning electron micrograph of a similar nanotube device, 
fabricated in the same batch as the measured device. The distance 
between source and drain electrodes is $\ell = 1 \,\upmu\text{m}$, the nanotube 
to gate distance is $d=210\un{nm}$. (b) Schematic drawing of the device, 
including a coplanar waveguide used as rf antenna and the cutting electrodes for 
nanotube transfer. (c) Dc current measurement showing Coulomb blockade 
oscillations. The absolute value of the current $I(\vg,\vsd)$ is plotted in 
logarithmic scale as a function of gate voltage \vg\ and bias voltage \vsd. (d) 
Data points: mechanical resonance frequencies of the two observed modes M1 and 
M2 of the device as a function of applied gate voltage \vg, see the 
Supplementary Material for the measurement data. Lines: polynomial fit as guide 
to the eye.
}
\end{figure}

Figure~\ref{fig-device}(a) shows a top view scanning electron micrograph of the 
measured device, Fig.~\ref{fig-device}(b) a larger scale sketch. A CNT is grown 
across the prongs of a quartz tuning fork and transferred onto lithographically 
defined source (S) and drain (D) titanium / gold contacts with a distance of 
$1\,\upmu\text{m}$ \cite{forktransfer}. In the region between the contacts, it 
is suspended over a gate electrode \cite{forktransfer, optomechanics}. The 
measurement was realized on a combined CNT -- coplanar waveguide resonator 
device similar to these used in \cite{optomechanics, stepwisefab, 
modelingomit}, however, here the additional ports only served as dc gate 
connection and MHz rf antenna for driving the nanotube into mechanical 
resonance.

At dilution refrigerator base temperature $T \simeq 13\un{mK}$, the CNT displays
typical quantum dot behaviour with moderately disordered Coulomb blockade
oscillations, see Fig. \ref{fig-device}(c). The gate lever arm is estimated as
$\alpha_\text{arm}=0.37$ and the charging energy as $E_C=9\un{meV}$. This leads to a gate
capacitance $\cg = 6.7\un{aF}$ and a total dot capacitance $\ctot = 18.1 
\un{aF}$. The CNT is set in motion by a MHz drive signal coupled into the 
device through the nearby coplanar waveguide. The resulting nanotube 
oscillation amplitude strongly increases once the drive signal hits the 
mechanical resonance frequency. The geometric CNT--gate capacitance \cg\ is 
modulated by the position of the nanotube with respect to the gate electrode. 
In a dc measurement as done here, the current averaged over these oscillations 
is detected. Its dependence on \cg\ allows us to trace the CNT resonance 
frequency as function of \vg\ \cite{highq, strongcoupling, prb-meerwaldt-2012, 
kondocharge}.

% \section{Electromechanical coupling}

\begin{figure}
\includegraphics{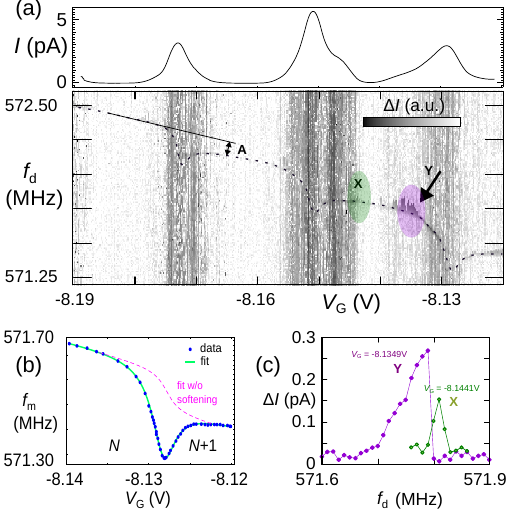}
\caption{\label{fig-softening}
Electromechanical softening of the nanotube spring constant due to single
electron tunneling. (a) Upper panel: Coulomb blockade oscillations of current 
$I(\vg)$ for a constant bias voltage $\vsd = 0.5\un{mV}$ and drive frequency 
$f_\text{d} = 572\un{MHz}$. Lower panel: Current $\Delta I(\vg, f_\text{d})$ as function of 
gate voltage \vg\ and drive frequency $f_\text{d}$, with a constant value for 
each frequency sweep subtracted to enhance details. Nominal drive power 
$P_\text{d} = -75\un{dBm}$ and bias voltage $\vsd = 0.5\un{mV}$. The dotted 
line is a guide to the eye for the mechanical resonance. 
(b) Data points: manually extracted 
resonance frequency points from (a) ($f_\text{m}(\vg)$) across one Coulomb peak 
displaying both charge-induced tension and electrostatic softening. The solid 
green line is a fit to the data using Eq.~(\ref{e1}); the dashed line omits the 
softening term (i.e., sets $S_4=0$) to illustrate the tension-induced tuning 
alone. (c) Line cuts from (a) in the regions marked there as X and Y, 
illustrating a barely resolved linear response peak (X) and a nonlinear Duffing 
response (Y) as a function of driving frequency $f_\text{d}$ in the current $\Delta 
I(\vg, f_\text{d}) = I(\vg, f_\text{d}) - \overline{I}(\vg)$.
}
\end{figure}

Effective coupling of motion to the charge of single tunneling electrons is a 
direct prerequisite for many interesting experiments \cite{optomechanics, 
nphys-wen-2020, prx-pistolesi-2021, nphys-samanta-2023}. In Coulomb blockade, 
where $E_C \gg k_B T$ and tunneling of electrons on single-electron tunneling 
current peaks occurs much faster than the mechanical oscillation $\Gamma 
\gg 2\pi f_\text{m}$, back action of the electronic tunneling causes an electrostatic 
softening leading to dips in resonance frequency $f_\text{m}$. These dips in the 
mechanical resonance frequency of the nanotube correspond to the Coulomb 
oscillations of conductance \cite{strongcoupling, prb-meerwaldt-2012}.

Figure~\ref{fig-softening}(a) shows a dc current measurement across three 
Coulomb oscillations, with a simultaneous rf driving signal of frequency $f_\text{d}$
applied on the nearby coplanar line. The measurement was performed at a low 
driving power of nominally $P_\text{d}=-75\un{dBm}$ at the chip socket (corresponding 
to a generator power of nominally $-30\un{dBm}$ and approximately $45\un{dB}$ 
input line attenuation), with the objective of observing linear oscillator 
response. The total dc current $I(\vg)$ at a fixed driving frequency is shown 
in the line plot of the upper panel; the lower panel displays the change in dc 
current $\Delta I(\vg, f_\text{d})$ as function of gate voltage and driving 
frequency, with a constant value for each frequency sweep subtracted. A finite 
bias $\vsd = 0.5\un{mV}$ was required to obtain a mechanical resonance signal. 
The resonance, here in the frequency range $571\un{MHz}\le f \le 
572.5\un{MHz}$, is traced by a dotted line as guide to the eye. Assuming a 
typical SW-CNT diameter of approximately 2 nm, the effective mass $m$ of the 
nanotube is estimated,  by multiplying the area of the CNT cylinder with the 
surface mass density of graphene \cite{nn-chen-2009, rmp-castroneto-2009} which 
results in $m = 4.8\cdot 10^{-21}\un{kg}$. Using the relation $f = 1/2 
\sqrt{T/m\ell}$, one obtains a tension $T=6.2\un{nN}$ on the CNT. Thus, the 
relatively large resonance frequency corresponds to a significant imprinted 
tension, most likely a result of our mechanical nanotube transfer technique 
\cite{forktransfer}. This places us in the so called high tension regime 
\cite{prb-meerwaldt-2012, rondac-lifshitz-2008, forktransfer, optomechanics}.

Near the Coulomb blockade oscillations, electrostatic softening of the nanotube 
spring constant $\alpha$ due to back action of tunneling electrons takes place, 
leading to distinct dips in resonance frequency \cite{strongcoupling, 
s-lassagne-2009}. The characteristic step in the resonance frequency 
corresponding to the addition of one elementary charge to the quantum dot is 
indicated with a double arrow \textbf{A} in Fig.~\ref{fig-softening}(a). The 
functional dependence of $f_\text{m}(\vg)$ can be modeled via the dependence of the 
resonance frequency on the time-averaged charge occupation $N(\vg)$ of the 
quantum dot with $\vg$, well-defined due to separation of time scales. The 
approach is given by \cite{modelingomit}
\begin{equation}
f(\vg)=S_1 + S_2 \vg + S_3 N(\vg) + S_4 \dfrac{\partial 
N}{\partial \vg} , \label{e1}
\end{equation}
where $S_1$ is a bare mechanical resonance frequency of the nanotube and $S_2$ 
the slope of the linear dependence of $f_\text{m}$ on $\vg$. $S_3$ describes the size 
of the step in $f_\text{m}$ across a single Coulomb oscillation, caused by an 
additional elementary charge on the quantum dot; cf.\ \textbf{A} in 
Fig.~\ref{fig-softening}(a). Last, $S_4=-\hbar g^2 C_\Sigma /{2\pi e \cg}$ 
captures the frequency dip or electrostatic softening that is a result of the 
nanotube's mechanical response to charge fluctuations; the magnitude of the 
frequency dip depends on the electromechanical coupling $g$, such that the fit 
allows us to extract the coupling parameter. Details of the derivation of $S_4$ 
are given in the supplementary material. For the bending mode considered in 
this work, the coupling particularly depends on how close the CNT is to the 
gate electrode \cite{prr-vigneau-2022, prl-micchi-2015, prb-micchi-2016}. Note 
for comparison that in \cite{arxiv-sevitz-2025} the working point is chosen 
such that there is approximately no large-scale charge dependence of the 
resonance frequency, in our notation corresponding to $S_2\simeq 0$ and $S_3 
\simeq 0$.

For modeling $N(\vg)$, we use a single Lorentz-broadened level coupled via 
tunnel barriers to Fermi distributions in the leads. This constructs
a fit function with free parameters $S_i$ as well as the common tunnel rate 
$\Gamma$ that follows the evolution of quantum dot average occupation across a 
single Coulomb oscillation \cite{strongcoupling, prb-meerwaldt-2012, 
optomechanics, modelingomit}. Fig.~\ref{fig-softening}(b) demonstrates this fit. 
Data points indicate the resonance frequencies extracted from a measurement 
as Fig.~\ref{fig-softening}(a). The solid green line is the fit using 
Eq.~\ref{e1}. The dashed line reproduces the fit result but with $S_4$ set to 
zero to illustrate the hypothetical resonance frequency in absence of 
electrostatic softening. The fit works surprisingly well despite the finite 
applied bias $\vsd=0.5\un{mV}$ \cite{prb-meerwaldt-2012}; this might be the 
case due to an asymmetry of the tunnel barriers, such that one edge of the 
SET region dominates the charge increase.

A nano-electromechanical system is said to be in ultrastrong electromechanical 
coupling when its coupling parameter $g/2\pi$ is larger than the mechanical 
resonance frequency $f_\text{m}$ \cite{nphys-samanta-2023, prr-vigneau-2022, 
prl-micchi-2015}. The coupling $g$ is related to the change of the 
electrochemical potential $\mu$ on the quantum dot via $\mu(z) \simeq \mu_0 
+\hbar g {z}/{z_\text{zpf}}$ \cite{prr-vigneau-2022}. Here, $z$ is the CNT 
deflection from equilibrium, $\mu_0$ is the quantum dot electrochemical 
potential at $z=0$, and $z_{\text{zpf}}= \sqrt{ \hbar / 2 m \omega_\text{m}}$ is the 
zero point fluctuation scale of the CNT as a mechanical oscillator. From the 
fit of Fig.~\ref{fig-softening}(b), a coupling $g/2\pi= 2.3\un{GHz}$ and a 
resulting coupling ratio $g/(2\pi f_\text{m})=4.0$ is extracted, confirming 
ultrastrong electromechanical coupling. This fulfils an important prerequisite 
for advanced nano-electromechanical experiments.

% \section{Nonlinear mechanics under strong drive}

With the typical high mechanical quality factor, the drive power range for a 
linear, harmonic oscillator response in a carbon nanotube is very small 
\cite{highq, nnano-moser-2014, apl-postma-2005}. Even in the carefully tuned 
measurement of Fig.~\ref{fig-softening}(a), for some values of the gate voltage 
\vg\ a nonlinear response with the typical Duffing curve can be observed 
\cite{highq, strongcoupling}, see Fig.~\ref{fig-softening}(c), while at other 
values the resonant signal is below detection range.
\begin{figure}[tb]
\includegraphics{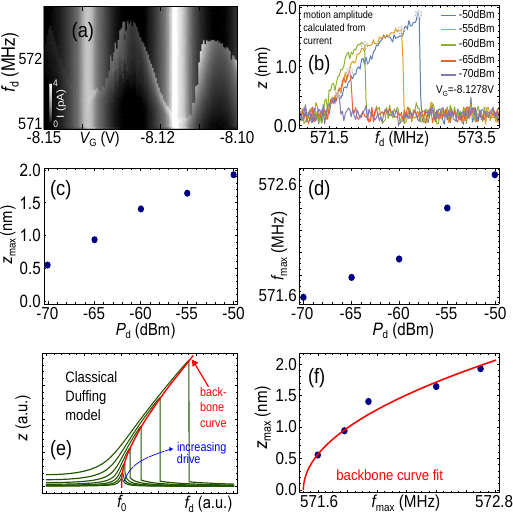}
\caption{\label{fig-duffing}
(a) Nonlinear nano-electromechanical response of the nanotube device at strong 
driving: measured dc current $I(\vg, f_\text{d})$ as function of gate voltage and 
drive frequency $f_\text{d}$ across two Coulomb oscillations, for a rf signal of $P_\text{d} 
= -55\un{dBm}$ at the chip holder; dc bias $\vsd \simeq 0$. The drive frequency 
$f_\text{d}$ is always swept in increasing direction. (b) Duffing response in 
the oscillation amplitude for different drive powers (corresponding to line 
cuts in measurements such as in (a)). Again, the drive frequency $f_\text{d}$ 
is always swept 
in increasing direction, $\vg=-8.1278\un{V}$. (c), (d) Jump-down points of motion 
amplitude (c) and frequency (d) from (b), plotted against applied drive power. 
(e) Schematic of the drive response of a classical Duffing oscillator.
(f) Example fit of a Duffing backbone curve to the jump-down points from the 
experiment (data from (c) and (d)).
}
\end{figure}
At millikelvin temperatures and increased power of the drive signal, the 
mechanical response of the nanotube rapidly becomes nonlinear for wide ranges 
of \vg\ \cite{highq}. This is illustrated in Fig.~\ref{fig-duffing}(a), 
plotting the dc current as function of gate voltage and drive frequency across 
two Coulomb blockade oscillations, now at $P_\text{d} = -55\un{dBm}$. The nonlinear 
artifacts are very strong, in particular in the Coulomb blockade regions where 
the high amplitude branch extends far upwards in frequency, completely 
overshadowing the gate voltage dependence of the linear response resonance 
frequency. Still, the softening effects due to single-electron tunneling at the 
conductance maxima remain visible.
 
In Fig.~\ref{fig-duffing}(b), line cuts at fixed gate voltage $\vg = -8.1278 
\un{V}$ for different drive powers and sweeps of increasing drive frequency 
$f_\text{d}$, demonstrates the evolving characteristic ``shark fin'' Duffing response 
of a nonlinear resonator with stiffening spring. Here, the oscillation 
amplitude $z$ derived from the resonant current is plotted; for details see 
\cite{highq, optomechanics} and the Supplementary Material. For a study of the 
nonlinear response, the jump-down points at the high-frequency edge of the 
bistability region are extracted for each of the Duffing response curves and 
plotted in Fig.~\ref{fig-duffing}(c) (maximum reached motion amplitude) and 
Fig.~\ref{fig-duffing}(d) (corresponding drive frequency).

The Duffing equation has been widely  employed in the modeling and analysis of
oscillating mechanical systems \cite{ajp-jones-2001, josav-brennan-2008,
josav-warminski-2009, c-clerc-2018, cinsans-ramlan-2016, josav-ghouli-2017,
rondac-lifshitz-2008, prl-catalini-2021, nl-kaisar-2022}. Its simplest form is 
given by \cite{book-duffing-1918, josav-brennan-2008, sr-wawrzynski-2022}
\begin{equation}
 m \ddot{z} + c\dot{z} + \alpha z + \beta z^3 = F \cos(2\pi f_\text{d} t).
\label{eq-duffing}
\end{equation}
Here, $m$ is the mass, $z$ the position of the oscillating object. $F$ is the
amplitude of the driving force with frequency $f_\text{d}$; $c$ describes the (linear)
damping, $\alpha$ the (harmonic oscillator) spring constant, and $\beta$ the
nonlinearity of the system. The polynomial $\alpha +\beta z^2$ represents the 
stiffness of the CNT spring with the linear response part controlled by 
$\alpha$ and nonlinear part controlled by the Duffing parameter $\beta$. For 
$\alpha>0$ and $\beta>0$, the spring constant of the oscillating system is said 
to be stiffening; conversely, for a softening system $\beta<0$ 
\cite{josav-brennan-2008, rondac-lifshitz-2008, sr-wawrzynski-2022}. 
Fig.~\ref{fig-duffing}(e) schematically displays the amplitude response curve 
of a stiffening Duffing system for increasing driving frequency. At strong 
driving, a Duffing system exhibits bistability; in the case plotted here, for 
increasing drive frequency, the amplitude follows the high-response branch over 
this region, and at its edge abruptly drops back to much lower value, resulting 
in the characteristic curve shape. For discussions of the case of gate voltage 
independent $c$ and $\beta$ in a carbon nanotube, see for example  
\cite{wang_visualizing_2021, arxiv-sevitz-2025}.

The drop-down point at the edge of the bistability region follows the so-called
backbone curve as function of drive power. A theoretical expression for this
curve can be derived from Eq.~\ref{eq-duffing} \cite{josav-brennan-2008, 
sr-wawrzynski-2022,-jordan-2023}; one obtains
\begin{equation}
f^2_\text{max}=\frac{1}{4\pi^2 m} \left(\frac{3}{4} \beta z^2_\text{max} + 
\alpha \right).
\label{eq-backbone}
\end{equation}  
Note that this expression does not directly contain the driving signal amplitude
anymore, only the jump-down frequency and maximum amplitude. In 
Fig.~\ref{fig-duffing}(f), this backbone curve is fitted to the jump-down
points in the Duffing response of the nanotube, i.e., the values from 
Fig.~\ref{fig-duffing}(c) and Fig.~\ref{fig-duffing}(d). Here we assume that the
frequency sweep is fast enough and spontaneous switching between the two stable
solutions of the Duffing equation is rare enough so the sweep always reaches 
the edge of the bistability region on the high-response branch. It is then 
possible to extract both the linear spring constant $\alpha$ and the Duffing 
parameter $\beta$; for the chosen fixed gate voltage of $\vg = -8.1278 \un{V}$, 
we obtain as fit parameters $\alpha = 6.189 \cdot 10^{-2} \, \text{kg} / 
\text{s}^2 $ and $\beta = 8.852\cdot10^{13}\,\text{kg}/\text{m}^2 \text{s}^2$. 

% \section*{Evolution of oscillator parameters}

\begin{figure}[tb]
\includegraphics{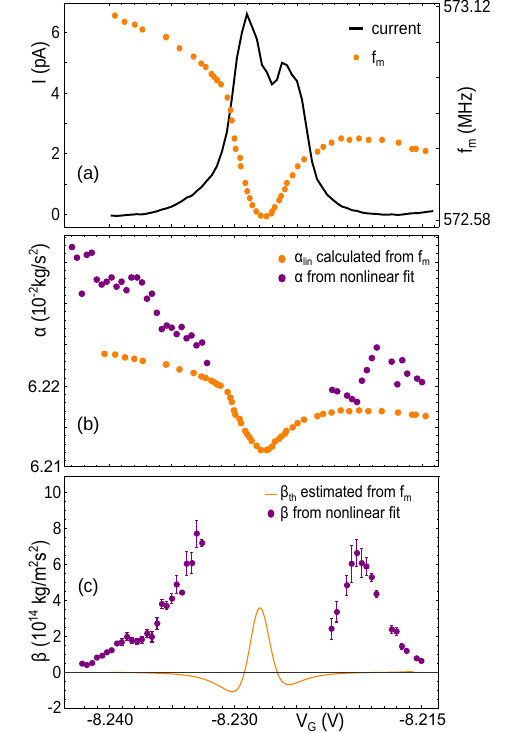}
\caption{\label{fig-devel}
Electronic and nano-electromechanical parameters at nonlinear driving as 
function of gate voltage \vg, for $\vsd\simeq 0$ (black / purple). Orange 
data points and lines stem from a different data set in linear mechanical 
response with $\vsd=0.5\un{mV}$ for comparison and have been corrected for a 
gate voltage offset. (a) dc current $I(\vg)$ and linear response resonance 
frequency $f_\text{m}(\vg)$. (b) Spring constant calculated from the linear response 
resonance frequency and extracted from Duffing response fits as in 
Fig.~\ref{fig-duffing}(f). (c) Duffing parameter $\beta(\vg)$ extracted from the 
Duffing response fits, and theoretical estimate of the SET Duffing parameter 
contribution $\beta_\text{th}(\vg)$ calculated from linear response, see the 
text.
}
\end{figure}

Figs.~\ref{fig-devel}(a-c) show the evolution of the electronic and
nano-electromechanical system parameters as function of gate voltage \vg\ 
across a Coulomb oscillation. Two separate data sets are combined. The black 
and purple lines and points show an evaluation of the response at nonlinear 
driving, with $\vsd\simeq 0$. The orange data points and lines in linear 
mechanical response stem from a measurement with $\vsd=0.5\un{mV}$, for better 
extraction of the linear resonance frequency, and have been corrected for a gate 
voltage offset.

Fig.~\ref{fig-devel}(a) plots the measured off-resonant dc current $I(\vg)$ 
across a Coulomb oscillation, in 
combination with the extracted resonance frequency $f_\text{m}(\vg)$ in linear 
response at weak driving. The linear response resonance frequency can be used 
to calculate the corresponding spring constant via $2\pi f_\text{m} = \sqrt{ 
\alpha_\text{lin} / m}$ with an estimated $m = 4.8\cdot 10^{-21}\un{kg}$ for 
this device. The resulting values $\alpha_\text{lin}(\vg)$ are plotted together 
with the $\alpha(\vg)$ from Duffing backbone fits in Fig.~\ref{fig-devel}(b) as 
consistency test. We observe reasonable agreement in magnitude and approximate 
functional behaviour. The Duffing evaluation leads to stronger scatter of the 
points as well as overall slightly higher values. The more complex evaluation 
procedure successfully identifies resonance peaks only at higher driving power, 
which for the hardening spring here indeed would lead to a higher identified 
value of the base resonance frequency and spring constant.

In Fig. \ref{fig-devel}(c), the Duffing parameter $\beta(\vg)$ is plotted. 
Additionally, the theoretically expected value $\beta_\text{th}(\vg)$ of the 
single-electron tunneling contribution is estimated from $\alpha_\text{lin} 
(\vg)$ using the relation \cite{strongcoupling}
\begin{equation}
 \beta_\text{th} = \frac{1}{6}
 \frac{\vg^2}{C_\text{G}^2}\,\left(\frac{\text{d}C_\text{G}}{\text{d}z}\right)^2
  \, \frac{\text{d}^2\alpha}{\text{d}\vg^2}.
\end{equation}
It effectively is the third derivative with respect to $z$ of the back action force on the quantum 
dot due to SET \cite{strongcoupling, prb-meerwaldt-2012}. While the absolute 
value of $\beta(\vg)$ behaves similar in both evaluations, the difference in 
sign is a striking deviation. Indeed, away from the center of the current peak, 
the SET contribution to the noninear behaviour should lead to a softening spring 
\cite{strongcoupling}, while our data displays stiffening spring behaviour.

Initial publications assumed that SET always dominates the nonlinear response 
\cite{strongcoupling, prb-meerwaldt-2012, optomechanics, modelingomit}, as 
compared to the non-SET related material nonlinearity which acts stiffening. In 
the meantime, also overall stiffening spring behaviour has been observed in 
experiments at low temperature \cite{willick_probing_2017, luo_coupling_2017, 
wang_visualizing_2021, arxiv-sevitz-2025}, with the typical explanation being 
that in these cases the material nonlinearity is the dominant contribution, 
possibly due to stronger driving. Furthermore, the material-induced nonlinear 
mechanical response is expected to grow with the tension in the nanotube. As a 
comparison, the nonlinear damping of an $l=840\un{nm}$ CNT under tension, $T = 
1.7\un{nN}$, in {\em absence} of Coulomb blockade has been studied and a 
nonlinear parameter $\beta = 6\cdot 10^{12} \,\text{kg}/\text{m}^2 
\text{s}^2$ extracted in Ref.~\cite{nnano-eichler-2011}, while our device is 
characterized by a larger value of $T = 6.2\un{nN}$ as mentioned above. This 
would however not explain the distinct gate voltage dependence of 
Fig.~\ref{fig-devel}(c), with the material constants independent of Coulomb 
blockade behaviour.

An interplay of both SET and material nonlinearity effects can be considered, 
however it provides no clear explanation of the growth of $\beta(\vg)$ near the 
SET current maximum. Even assuming a straightforward addition of the 
contributions, the differences in feature width between observation and theory 
curve as well as the contributions of different sign indicate complex 
behaviour. 

A switch in sign of the SET-induced nonlinearity is expected at the flanks of 
the Coulomb oscillation, indicating a transition from a softening to a 
hardening spring constant SET contribution or reverse, as shown in 
Refs.~\cite{strongcoupling, prb-meerwaldt-2012}. In our data, no Duffing curve 
can be identified at the current peaks, see Fig. \ref{fig-devel}(c), leading to 
missing data points in the evaluation. This was also observed for several other 
Coulomb oscillations not shown here and may be related to nonlinear damping 
which should scale with nonlinearity \cite{nnano-eichler-2011}. As a result, 
further discussion of the gate voltage dependence of $\beta(\vg)$ enters the 
realm of speculation, and further experiments are required to enable a deeper 
understanding of this phenomenon.

% \section*{Summary and conclusion}

In conclusion, we have studied the nonlinear mechanics of a suspended carbon
nanotube quantum dot at dilution refrigerator base temperature. Motion and 
electron tunneling show ultrastrong coupling, at a high mechanical resonance 
frequency $f\sim 570\un{MHz}$ due to fabrication-imprinted tension. When 
driving, we obtain the characteristic frequency response curve of a Duffing 
oscillator. Fitting the Duffing backbone curve enables us to extract both 
linear stiffness $\alpha$ and Duffing parameter $\beta$ as a function of gate 
voltage. The result demonstrates a rich interplay of Coulomb blockade, material 
properties and the nano-electromechanical system parameters. While displaying a 
clear gate voltage dependence and thus indicating SET-induced nonlinearity 
effects, our device shows throughout a stiffening spring behaviour, in a 
puzzling deviation from expectations.

Characterization and control of the nonlinearity of a vibrating CNT with strong 
coupling between mechanical motion and single-electron tunneling is an 
important prerequisite for quantum device applications. Recently, a hybrid 
nano-electromechanical qubit based on a double quantum dot within a suspended 
CNT has been proposed \cite{prx-pistolesi-2021}; the outstanding mechanical 
quality factors of carbon nanotubes at cryogenic temperature promise long 
storage times for quantum information. Correspondingly, in particular for 
addressing and controlling the mechanical quantum regime, the deeper 
understanding of the interplay of linear (i.e. resonator- or oscillator-like) 
and nonlinear (i.e. qubit-like) nano-electromechanical quantum phenomena in a 
carbon nanotube will be of central interest.

\begin{acknowledgments}
The authors acknowledge financial support by the Deutsche 
Forschungsgemeinschaft via grants Hu 1808/4 (project id 438638106), Hu 
1808/5 (project id 438640202), and SFB 1277 (project id 314695032). We would 
like to thank P.~Hakonen for insightful discussions, and Ch.~Strunk and D.~Weiss for the use of experimental facilities. The 
measurement data was recorded using Lab::Measurement \cite{labmeasurement}.
\end{acknowledgments}

\vspace*{0.2cm}
\section*{Author contributions}

A.~K.~H. and A.~N.~L. conceived and designed the experiment. The coplanar 
waveguide device was developed and fabricated by K.~B. and A.~N.~L., with help 
and advice from N.~H. Nanotube growth and transfer were established 
and optimized by F.~Ö., F.~S., and A.~N.~L. The low temperature measurements 
were performed by A.~N.~L.; data evaluation and manuscript writing were done 
jointly by A.~N.~L. and A.~K.~H. The project was supervised by A.~K.~H.

\bibliography{duffing}

\begin{thebibliography}{46}%
\makeatletter
\providecommand \@ifxundefined [1]{%
 \@ifx{#1\undefined}
}%
\providecommand \@ifnum [1]{%
 \ifnum #1\expandafter \@firstoftwo
 \else \expandafter \@secondoftwo
 \fi
}%
\providecommand \@ifx [1]{%
 \ifx #1\expandafter \@firstoftwo
 \else \expandafter \@secondoftwo
 \fi
}%
\providecommand \natexlab [1]{#1}%
\providecommand \enquote  [1]{``#1''}%
\providecommand \bibnamefont  [1]{#1}%
\providecommand \bibfnamefont [1]{#1}%
\providecommand \citenamefont [1]{#1}%
\providecommand \href@noop [0]{\@secondoftwo}%
\providecommand \href [0]{\begingroup \@sanitize@url \@href}%
\providecommand \@href[1]{\@@startlink{#1}\@@href}%
\providecommand \@@href[1]{\endgroup#1\@@endlink}%
\providecommand \@sanitize@url [0]{\catcode `\\12\catcode `\$12\catcode
  `\&12\catcode `\#12\catcode `\^12\catcode `\_12\catcode `\%12\relax}%
\providecommand \@@startlink[1]{}%
\providecommand \@@endlink[0]{}%
\providecommand \url  [0]{\begingroup\@sanitize@url \@url }%
\providecommand \@url [1]{\endgroup\@href {#1}{\urlprefix }}%
\providecommand \urlprefix  [0]{URL }%
\providecommand \Eprint [0]{\href }%
\providecommand \doibase [0]{https://doi.org/}%
\providecommand \selectlanguage [0]{\@gobble}%
\providecommand \bibinfo  [0]{\@secondoftwo}%
\providecommand \bibfield  [0]{\@secondoftwo}%
\providecommand \translation [1]{[#1]}%
\providecommand \BibitemOpen [0]{}%
\providecommand \bibitemStop [0]{}%
\providecommand \bibitemNoStop [0]{.\EOS\space}%
\providecommand \EOS [0]{\spacefactor3000\relax}%
\providecommand \BibitemShut  [1]{\csname bibitem#1\endcsname}%
\let\auto@bib@innerbib\@empty
%</preamble>
\bibitem [{\citenamefont {Sazonova}\ \emph {et~al.}(2004)\citenamefont
  {Sazonova}, \citenamefont {Yaish}, \citenamefont {{\"U}st{\"u}nel},
  \citenamefont {Roundy}, \citenamefont {Arias},\ and\ \citenamefont
  {McEuen}}]{nature-sazonova-2004}%
  \BibitemOpen
  \bibfield  {author} {\bibinfo {author} {\bibfnamefont {V.}~\bibnamefont
  {Sazonova}}, \bibinfo {author} {\bibfnamefont {Y.}~\bibnamefont {Yaish}},
  \bibinfo {author} {\bibfnamefont {H.}~\bibnamefont {{\"U}st{\"u}nel}},
  \bibinfo {author} {\bibfnamefont {D.}~\bibnamefont {Roundy}}, \bibinfo
  {author} {\bibfnamefont {T.~A.}\ \bibnamefont {Arias}},\ and\ \bibinfo
  {author} {\bibfnamefont {P.~L.}\ \bibnamefont {McEuen}},\ }\bibfield  {title}
  {\bibinfo {title} {A tunable carbon nanotube electromechanical oscillator},\
  }\href {https://doi.org/10.1038/nature02905} {\bibfield  {journal} {\bibinfo
  {journal} {Nature}\ }\textbf {\bibinfo {volume} {431}},\ \bibinfo {pages}
  {284} (\bibinfo {year} {2004})}\BibitemShut {NoStop}%
\bibitem [{\citenamefont {H{\"u}ttel}\ \emph {et~al.}(2009)\citenamefont
  {H{\"u}ttel}, \citenamefont {Steele}, \citenamefont {Witkamp}, \citenamefont
  {Poot}, \citenamefont {Kouwenhoven},\ and\ \citenamefont {{van der
  Zant}}}]{highq}%
  \BibitemOpen
  \bibfield  {author} {\bibinfo {author} {\bibfnamefont {A.~K.}\ \bibnamefont
  {H{\"u}ttel}}, \bibinfo {author} {\bibfnamefont {G.~A.}\ \bibnamefont
  {Steele}}, \bibinfo {author} {\bibfnamefont {B.}~\bibnamefont {Witkamp}},
  \bibinfo {author} {\bibfnamefont {M.}~\bibnamefont {Poot}}, \bibinfo {author}
  {\bibfnamefont {L.~P.}\ \bibnamefont {Kouwenhoven}},\ and\ \bibinfo {author}
  {\bibfnamefont {H.~S.~J.}\ \bibnamefont {{van der Zant}}},\ }\bibfield
  {title} {\bibinfo {title} {Carbon nanotubes as ultrahigh quality factor
  mechanical resonators},\ }\href {https://doi.org/10.1021/nl900612h}
  {\bibfield  {journal} {\bibinfo  {journal} {Nano Letters}\ }\textbf {\bibinfo
  {volume} {9}},\ \bibinfo {pages} {2547} (\bibinfo {year} {2009})}\BibitemShut
  {NoStop}%
\bibitem [{\citenamefont {Moser}\ \emph {et~al.}(2014)\citenamefont {Moser},
  \citenamefont {Eichler}, \citenamefont {G{\"u}ttinger}, \citenamefont
  {Dykman},\ and\ \citenamefont {Bachtold}}]{nnano-moser-2014}%
  \BibitemOpen
  \bibfield  {author} {\bibinfo {author} {\bibfnamefont {J.}~\bibnamefont
  {Moser}}, \bibinfo {author} {\bibfnamefont {A.}~\bibnamefont {Eichler}},
  \bibinfo {author} {\bibfnamefont {J.}~\bibnamefont {G{\"u}ttinger}}, \bibinfo
  {author} {\bibfnamefont {M.~I.}\ \bibnamefont {Dykman}},\ and\ \bibinfo
  {author} {\bibfnamefont {A.}~\bibnamefont {Bachtold}},\ }\bibfield  {title}
  {\bibinfo {title} {Nanotube mechanical resonators with quality factors of up
  to 5 million},\ }\href {https://doi.org/10.1038/nnano.2014.234} {\bibfield
  {journal} {\bibinfo  {journal} {Nature Nanotechnology}\ }\textbf {\bibinfo
  {volume} {9}},\ \bibinfo {pages} {1007} (\bibinfo {year} {2014})}\BibitemShut
  {NoStop}%
\bibitem [{\citenamefont {Laird}\ \emph {et~al.}(2015)\citenamefont {Laird},
  \citenamefont {Kuemmeth}, \citenamefont {Steele}, \citenamefont
  {{Grove-Rasmussen}}, \citenamefont {Nyg{\aa}rd}, \citenamefont {Flensberg},\
  and\ \citenamefont {Kouwenhoven}}]{rmp-laird-2015}%
  \BibitemOpen
  \bibfield  {author} {\bibinfo {author} {\bibfnamefont {E.~A.}\ \bibnamefont
  {Laird}}, \bibinfo {author} {\bibfnamefont {F.}~\bibnamefont {Kuemmeth}},
  \bibinfo {author} {\bibfnamefont {G.~A.}\ \bibnamefont {Steele}}, \bibinfo
  {author} {\bibfnamefont {K.}~\bibnamefont {{Grove-Rasmussen}}}, \bibinfo
  {author} {\bibfnamefont {J.}~\bibnamefont {Nyg{\aa}rd}}, \bibinfo {author}
  {\bibfnamefont {K.}~\bibnamefont {Flensberg}},\ and\ \bibinfo {author}
  {\bibfnamefont {L.~P.}\ \bibnamefont {Kouwenhoven}},\ }\bibfield  {title}
  {\bibinfo {title} {Quantum transport in carbon nanotubes},\ }\href
  {https://doi.org/10.1103/RevModPhys.87.703} {\bibfield  {journal} {\bibinfo
  {journal} {Reviews of Modern Physics}\ }\textbf {\bibinfo {volume} {87}},\
  \bibinfo {pages} {703} (\bibinfo {year} {2015})}\BibitemShut {NoStop}%
\bibitem [{\citenamefont {Bachtold}\ \emph {et~al.}(2022)\citenamefont
  {Bachtold}, \citenamefont {Moser},\ and\ \citenamefont
  {Dykman}}]{rmp-bachtold-2022}%
  \BibitemOpen
  \bibfield  {author} {\bibinfo {author} {\bibfnamefont {A.}~\bibnamefont
  {Bachtold}}, \bibinfo {author} {\bibfnamefont {J.}~\bibnamefont {Moser}},\
  and\ \bibinfo {author} {\bibfnamefont {M.~I.}\ \bibnamefont {Dykman}},\
  }\bibfield  {title} {\bibinfo {title} {Mesoscopic physics of nanomechanical
  systems},\ }\href {https://doi.org/10.1103/RevModPhys.94.045005} {\bibfield
  {journal} {\bibinfo  {journal} {Reviews of Modern Physics}\ }\textbf
  {\bibinfo {volume} {94}},\ \bibinfo {pages} {045005} (\bibinfo {year}
  {2022})}\BibitemShut {NoStop}%
\bibitem [{\citenamefont {Sevitz}\ \emph {et~al.}(2025)\citenamefont {Sevitz},
  \citenamefont {Aggarwal}, \citenamefont {{Tabanera-Bravo}}, \citenamefont
  {Monsel}, \citenamefont {Vigneau}, \citenamefont {Fedele}, \citenamefont
  {Dunlop}, \citenamefont {Parrondo}, \citenamefont {Milburn}, \citenamefont
  {Anders}, \citenamefont {Ares},\ and\ \citenamefont
  {Cerisola}}]{arxiv-sevitz-2025}%
  \BibitemOpen
  \bibfield  {author} {\bibinfo {author} {\bibfnamefont {S.}~\bibnamefont
  {Sevitz}}, \bibinfo {author} {\bibfnamefont {K.}~\bibnamefont {Aggarwal}},
  \bibinfo {author} {\bibfnamefont {J.}~\bibnamefont {{Tabanera-Bravo}}},
  \bibinfo {author} {\bibfnamefont {J.}~\bibnamefont {Monsel}}, \bibinfo
  {author} {\bibfnamefont {F.}~\bibnamefont {Vigneau}}, \bibinfo {author}
  {\bibfnamefont {F.}~\bibnamefont {Fedele}}, \bibinfo {author} {\bibfnamefont
  {J.}~\bibnamefont {Dunlop}}, \bibinfo {author} {\bibfnamefont {J.~M.~R.}\
  \bibnamefont {Parrondo}}, \bibinfo {author} {\bibfnamefont {G.~J.}\
  \bibnamefont {Milburn}}, \bibinfo {author} {\bibfnamefont {J.}~\bibnamefont
  {Anders}}, \bibinfo {author} {\bibfnamefont {N.}~\bibnamefont {Ares}},\ and\
  \bibinfo {author} {\bibfnamefont {F.}~\bibnamefont {Cerisola}},\ }\href@noop
  {} {\bibinfo {title} {Sources of nonlinearity in the response of a driven
  nano-electromechanical resonator}} (\bibinfo {year} {2025}),\ \Eprint
  {https://arxiv.org/abs/2509.12830} {arXiv:2509.12830 [cond-mat.mes-hall]}
  \BibitemShut {NoStop}%
\bibitem [{\citenamefont {Lassagne}\ \emph {et~al.}(2008)\citenamefont
  {Lassagne}, \citenamefont {{Garcia-Sanchez}}, \citenamefont {Aguasca},\ and\
  \citenamefont {Bachtold}}]{nl-lassagne-2008}%
  \BibitemOpen
  \bibfield  {author} {\bibinfo {author} {\bibfnamefont {B.}~\bibnamefont
  {Lassagne}}, \bibinfo {author} {\bibfnamefont {D.}~\bibnamefont
  {{Garcia-Sanchez}}}, \bibinfo {author} {\bibfnamefont {A.}~\bibnamefont
  {Aguasca}},\ and\ \bibinfo {author} {\bibfnamefont {A.}~\bibnamefont
  {Bachtold}},\ }\bibfield  {title} {\bibinfo {title} {Ultrasensitive {{Mass
  Sensing}} with a {{Nanotube Electromechanical Resonator}}},\ }\href
  {https://doi.org/10.1021/nl801982v} {\bibfield  {journal} {\bibinfo
  {journal} {Nano Letters}\ }\textbf {\bibinfo {volume} {8}},\ \bibinfo {pages}
  {3735} (\bibinfo {year} {2008})}\BibitemShut {NoStop}%
\bibitem [{\citenamefont {H{\"a}kkinen}\ \emph {et~al.}(2015)\citenamefont
  {H{\"a}kkinen}, \citenamefont {Isacsson}, \citenamefont {Savin},
  \citenamefont {Sulkko},\ and\ \citenamefont {Hakonen}}]{nl-hakkinen-2015}%
  \BibitemOpen
  \bibfield  {author} {\bibinfo {author} {\bibfnamefont {P.}~\bibnamefont
  {H{\"a}kkinen}}, \bibinfo {author} {\bibfnamefont {A.}~\bibnamefont
  {Isacsson}}, \bibinfo {author} {\bibfnamefont {A.}~\bibnamefont {Savin}},
  \bibinfo {author} {\bibfnamefont {J.}~\bibnamefont {Sulkko}},\ and\ \bibinfo
  {author} {\bibfnamefont {P.}~\bibnamefont {Hakonen}},\ }\bibfield  {title}
  {\bibinfo {title} {Charge sensitivity enhancement via mechanical oscillation
  in suspended carbon nanotube devices},\ }\href
  {https://doi.org/10.1021/nl504282s} {\bibfield  {journal} {\bibinfo
  {journal} {Nano Letters}\ }\textbf {\bibinfo {volume} {15}},\ \bibinfo
  {pages} {1667} (\bibinfo {year} {2015})}\BibitemShut {NoStop}%
\bibitem [{\citenamefont {{de Bonis}}\ \emph {et~al.}(2018)\citenamefont {{de
  Bonis}}, \citenamefont {Urgell}, \citenamefont {Yang}, \citenamefont
  {Samanta}, \citenamefont {Noury}, \citenamefont {{Vergara-Cruz}},
  \citenamefont {Dong}, \citenamefont {Jin},\ and\ \citenamefont
  {Bachtold}}]{nl-debonis-2018}%
  \BibitemOpen
  \bibfield  {author} {\bibinfo {author} {\bibfnamefont {S.~L.}\ \bibnamefont
  {{de Bonis}}}, \bibinfo {author} {\bibfnamefont {C.}~\bibnamefont {Urgell}},
  \bibinfo {author} {\bibfnamefont {W.}~\bibnamefont {Yang}}, \bibinfo {author}
  {\bibfnamefont {C.}~\bibnamefont {Samanta}}, \bibinfo {author} {\bibfnamefont
  {A.}~\bibnamefont {Noury}}, \bibinfo {author} {\bibfnamefont
  {J.}~\bibnamefont {{Vergara-Cruz}}}, \bibinfo {author} {\bibfnamefont
  {Q.}~\bibnamefont {Dong}}, \bibinfo {author} {\bibfnamefont {Y.}~\bibnamefont
  {Jin}},\ and\ \bibinfo {author} {\bibfnamefont {A.}~\bibnamefont
  {Bachtold}},\ }\bibfield  {title} {\bibinfo {title} {Ultrasensitive
  {{Displacement Noise Measurement}} of {{Carbon Nanotube Mechanical
  Resonators}}},\ }\href {https://doi.org/10.1021/acs.nanolett.8b02437}
  {\bibfield  {journal} {\bibinfo  {journal} {Nano Letters}\ }\textbf {\bibinfo
  {volume} {18}},\ \bibinfo {pages} {5324} (\bibinfo {year}
  {2018})}\BibitemShut {NoStop}%
\bibitem [{\citenamefont {Steele}\ \emph {et~al.}(2009)\citenamefont {Steele},
  \citenamefont {H{\"u}ttel}, \citenamefont {Witkamp}, \citenamefont {Poot},
  \citenamefont {Meerwaldt}, \citenamefont {Kouwenhoven},\ and\ \citenamefont
  {{van der Zant}}}]{strongcoupling}%
  \BibitemOpen
  \bibfield  {author} {\bibinfo {author} {\bibfnamefont {G.~A.}\ \bibnamefont
  {Steele}}, \bibinfo {author} {\bibfnamefont {A.~K.}\ \bibnamefont
  {H{\"u}ttel}}, \bibinfo {author} {\bibfnamefont {B.}~\bibnamefont {Witkamp}},
  \bibinfo {author} {\bibfnamefont {M.}~\bibnamefont {Poot}}, \bibinfo {author}
  {\bibfnamefont {H.~B.}\ \bibnamefont {Meerwaldt}}, \bibinfo {author}
  {\bibfnamefont {L.~P.}\ \bibnamefont {Kouwenhoven}},\ and\ \bibinfo {author}
  {\bibfnamefont {H.~S.~J.}\ \bibnamefont {{van der Zant}}},\ }\bibfield
  {title} {\bibinfo {title} {Strong coupling between single-electron tunneling
  and nanomechanical motion},\ }\href {https://doi.org/10.1126/science.1176076}
  {\bibfield  {journal} {\bibinfo  {journal} {Science}\ }\textbf {\bibinfo
  {volume} {325}},\ \bibinfo {pages} {1103} (\bibinfo {year}
  {2009})}\BibitemShut {NoStop}%
\bibitem [{\citenamefont {Lassagne}\ \emph {et~al.}(2009)\citenamefont
  {Lassagne}, \citenamefont {Tarakanov}, \citenamefont {Kinaret}, \citenamefont
  {{Garcia-Sanchez}},\ and\ \citenamefont {Bachtold}}]{s-lassagne-2009}%
  \BibitemOpen
  \bibfield  {author} {\bibinfo {author} {\bibfnamefont {B.}~\bibnamefont
  {Lassagne}}, \bibinfo {author} {\bibfnamefont {Y.}~\bibnamefont {Tarakanov}},
  \bibinfo {author} {\bibfnamefont {J.}~\bibnamefont {Kinaret}}, \bibinfo
  {author} {\bibfnamefont {D.}~\bibnamefont {{Garcia-Sanchez}}},\ and\ \bibinfo
  {author} {\bibfnamefont {A.}~\bibnamefont {Bachtold}},\ }\bibfield  {title}
  {\bibinfo {title} {Coupling {{Mechanics}} to {{Charge Transport}} in {{Carbon
  Nanotube Mechanical Resonators}}},\ }\href
  {https://doi.org/10.1126/science.1174290} {\bibfield  {journal} {\bibinfo
  {journal} {Science}\ }\textbf {\bibinfo {volume} {325}},\ \bibinfo {pages}
  {1107} (\bibinfo {year} {2009})}\BibitemShut {NoStop}%
\bibitem [{\citenamefont {Meerwaldt}\ \emph {et~al.}(2012)\citenamefont
  {Meerwaldt}, \citenamefont {Labadze}, \citenamefont {Schneider},
  \citenamefont {Taspinar}, \citenamefont {Blanter}, \citenamefont {{van der
  Zant}},\ and\ \citenamefont {Steele}}]{prb-meerwaldt-2012}%
  \BibitemOpen
  \bibfield  {author} {\bibinfo {author} {\bibfnamefont {H.~B.}\ \bibnamefont
  {Meerwaldt}}, \bibinfo {author} {\bibfnamefont {G.}~\bibnamefont {Labadze}},
  \bibinfo {author} {\bibfnamefont {B.~H.}\ \bibnamefont {Schneider}}, \bibinfo
  {author} {\bibfnamefont {A.}~\bibnamefont {Taspinar}}, \bibinfo {author}
  {\bibfnamefont {{\relax Ya}.~M.}\ \bibnamefont {Blanter}}, \bibinfo {author}
  {\bibfnamefont {H.~S.~J.}\ \bibnamefont {{van der Zant}}},\ and\ \bibinfo
  {author} {\bibfnamefont {G.~A.}\ \bibnamefont {Steele}},\ }\bibfield  {title}
  {\bibinfo {title} {Probing the charge of a quantum dot with a nanomechanical
  resonator},\ }\href {https://doi.org/10.1103/PhysRevB.86.115454} {\bibfield
  {journal} {\bibinfo  {journal} {Physical Review B}\ }\textbf {\bibinfo
  {volume} {86}},\ \bibinfo {pages} {115454} (\bibinfo {year}
  {2012})}\BibitemShut {NoStop}%
\bibitem [{\citenamefont {Schmid}\ \emph {et~al.}(2015)\citenamefont {Schmid},
  \citenamefont {Stiller}, \citenamefont {Strunk},\ and\ \citenamefont
  {H{\"u}ttel}}]{heliumdamping}%
  \BibitemOpen
  \bibfield  {author} {\bibinfo {author} {\bibfnamefont {D.~R.}\ \bibnamefont
  {Schmid}}, \bibinfo {author} {\bibfnamefont {P.~L.}\ \bibnamefont {Stiller}},
  \bibinfo {author} {\bibfnamefont {{\relax Ch}.}~\bibnamefont {Strunk}},\ and\
  \bibinfo {author} {\bibfnamefont {A.~K.}\ \bibnamefont {H{\"u}ttel}},\
  }\bibfield  {title} {\bibinfo {title} {Liquid-induced damping of mechanical
  feedback effects in single electron tunneling through a suspended carbon
  nanotube},\ }\href {https://doi.org/10.1063/1.4931775} {\bibfield  {journal}
  {\bibinfo  {journal} {Applied Physics Letters}\ }\textbf {\bibinfo {volume}
  {107}},\ \bibinfo {pages} {123110} (\bibinfo {year} {2015})}\BibitemShut
  {NoStop}%
\bibitem [{\citenamefont {Deng}\ \emph {et~al.}(2016)\citenamefont {Deng},
  \citenamefont {Zhu}, \citenamefont {Wang}, \citenamefont {Zou}, \citenamefont
  {Wang}, \citenamefont {Li}, \citenamefont {Cao}, \citenamefont {Liu},
  \citenamefont {Li}, \citenamefont {Xiao}, \citenamefont {Guo}, \citenamefont
  {Jiang}, \citenamefont {Dai},\ and\ \citenamefont
  {Guo}}]{deng_strongly_2016}%
  \BibitemOpen
  \bibfield  {author} {\bibinfo {author} {\bibfnamefont {G.-W.}\ \bibnamefont
  {Deng}}, \bibinfo {author} {\bibfnamefont {D.}~\bibnamefont {Zhu}}, \bibinfo
  {author} {\bibfnamefont {X.-H.}\ \bibnamefont {Wang}}, \bibinfo {author}
  {\bibfnamefont {C.-L.}\ \bibnamefont {Zou}}, \bibinfo {author} {\bibfnamefont
  {J.-T.}\ \bibnamefont {Wang}}, \bibinfo {author} {\bibfnamefont {H.-O.}\
  \bibnamefont {Li}}, \bibinfo {author} {\bibfnamefont {G.}~\bibnamefont
  {Cao}}, \bibinfo {author} {\bibfnamefont {D.}~\bibnamefont {Liu}}, \bibinfo
  {author} {\bibfnamefont {Y.}~\bibnamefont {Li}}, \bibinfo {author}
  {\bibfnamefont {M.}~\bibnamefont {Xiao}}, \bibinfo {author} {\bibfnamefont
  {G.-C.}\ \bibnamefont {Guo}}, \bibinfo {author} {\bibfnamefont {K.-L.}\
  \bibnamefont {Jiang}}, \bibinfo {author} {\bibfnamefont {X.-C.}\ \bibnamefont
  {Dai}},\ and\ \bibinfo {author} {\bibfnamefont {G.-P.}\ \bibnamefont {Guo}},\
  }\bibfield  {title} {\bibinfo {title} {Strongly coupled nanotube
  electromechanical resonators},\ }\href
  {https://doi.org/10.1021/acs.nanolett.6b01875} {\bibfield  {journal}
  {\bibinfo  {journal} {Nano Letters}\ }\textbf {\bibinfo {volume} {16}},\
  \bibinfo {pages} {5456} (\bibinfo {year} {2016})}\BibitemShut {NoStop}%
\bibitem [{\citenamefont {Urgell}\ \emph {et~al.}(2020)\citenamefont {Urgell},
  \citenamefont {Yang}, \citenamefont {De~Bonis}, \citenamefont {Samanta},
  \citenamefont {Esplandiu}, \citenamefont {Dong}, \citenamefont {Jin},\ and\
  \citenamefont {Bachtold}}]{nphys-urgell-2020}%
  \BibitemOpen
  \bibfield  {author} {\bibinfo {author} {\bibfnamefont {C.}~\bibnamefont
  {Urgell}}, \bibinfo {author} {\bibfnamefont {W.}~\bibnamefont {Yang}},
  \bibinfo {author} {\bibfnamefont {S.~L.}\ \bibnamefont {De~Bonis}}, \bibinfo
  {author} {\bibfnamefont {C.}~\bibnamefont {Samanta}}, \bibinfo {author}
  {\bibfnamefont {M.~J.}\ \bibnamefont {Esplandiu}}, \bibinfo {author}
  {\bibfnamefont {Q.}~\bibnamefont {Dong}}, \bibinfo {author} {\bibfnamefont
  {Y.}~\bibnamefont {Jin}},\ and\ \bibinfo {author} {\bibfnamefont
  {A.}~\bibnamefont {Bachtold}},\ }\bibfield  {title} {\bibinfo {title}
  {Cooling and self-oscillation in a nanotube electromechanical resonator},\
  }\href {https://doi.org/10.1038/s41567-019-0682-6} {\bibfield  {journal}
  {\bibinfo  {journal} {Nature Physics}\ }\textbf {\bibinfo {volume} {16}},\
  \bibinfo {pages} {32} (\bibinfo {year} {2020})}\BibitemShut {NoStop}%
\bibitem [{\citenamefont {Blien}\ \emph {et~al.}(2020)\citenamefont {Blien},
  \citenamefont {Steger}, \citenamefont {H{\"u}ttner}, \citenamefont {Graaf},\
  and\ \citenamefont {H{\"u}ttel}}]{optomechanics}%
  \BibitemOpen
  \bibfield  {author} {\bibinfo {author} {\bibfnamefont {S.}~\bibnamefont
  {Blien}}, \bibinfo {author} {\bibfnamefont {P.}~\bibnamefont {Steger}},
  \bibinfo {author} {\bibfnamefont {N.}~\bibnamefont {H{\"u}ttner}}, \bibinfo
  {author} {\bibfnamefont {R.}~\bibnamefont {Graaf}},\ and\ \bibinfo {author}
  {\bibfnamefont {A.~K.}\ \bibnamefont {H{\"u}ttel}},\ }\bibfield  {title}
  {\bibinfo {title} {Quantum capacitance mediated carbon nanotube
  optomechanics},\ }\href {https://doi.org/10.1038/s41467-020-15433-3}
  {\bibfield  {journal} {\bibinfo  {journal} {Nature Communications}\ }\textbf
  {\bibinfo {volume} {11}},\ \bibinfo {pages} {1363} (\bibinfo {year}
  {2020})}\BibitemShut {NoStop}%
\bibitem [{\citenamefont {Samanta}\ \emph {et~al.}(2023)\citenamefont
  {Samanta}, \citenamefont {De~Bonis}, \citenamefont {M{\o}ller}, \citenamefont
  {{Tormo-Queralt}}, \citenamefont {Yang}, \citenamefont {Urgell},
  \citenamefont {Stamenic}, \citenamefont {Thibeault}, \citenamefont {Jin},
  \citenamefont {Czaplewski}, \citenamefont {Pistolesi},\ and\ \citenamefont
  {Bachtold}}]{nphys-samanta-2023}%
  \BibitemOpen
  \bibfield  {author} {\bibinfo {author} {\bibfnamefont {C.}~\bibnamefont
  {Samanta}}, \bibinfo {author} {\bibfnamefont {S.~L.}\ \bibnamefont
  {De~Bonis}}, \bibinfo {author} {\bibfnamefont {C.~B.}\ \bibnamefont
  {M{\o}ller}}, \bibinfo {author} {\bibfnamefont {R.}~\bibnamefont
  {{Tormo-Queralt}}}, \bibinfo {author} {\bibfnamefont {W.}~\bibnamefont
  {Yang}}, \bibinfo {author} {\bibfnamefont {C.}~\bibnamefont {Urgell}},
  \bibinfo {author} {\bibfnamefont {B.}~\bibnamefont {Stamenic}}, \bibinfo
  {author} {\bibfnamefont {B.}~\bibnamefont {Thibeault}}, \bibinfo {author}
  {\bibfnamefont {Y.}~\bibnamefont {Jin}}, \bibinfo {author} {\bibfnamefont
  {D.~A.}\ \bibnamefont {Czaplewski}}, \bibinfo {author} {\bibfnamefont
  {F.}~\bibnamefont {Pistolesi}},\ and\ \bibinfo {author} {\bibfnamefont
  {A.}~\bibnamefont {Bachtold}},\ }\bibfield  {title} {\bibinfo {title}
  {Nonlinear nanomechanical resonators approaching the quantum ground state},\
  }\href {https://doi.org/10.1038/s41567-023-02065-9} {\bibfield  {journal}
  {\bibinfo  {journal} {Nature Physics}\ }\textbf {\bibinfo {volume} {19}},\
  \bibinfo {pages} {1340} (\bibinfo {year} {2023})}\BibitemShut {NoStop}%
\bibitem [{\citenamefont {Pistolesi}\ \emph {et~al.}(2021)\citenamefont
  {Pistolesi}, \citenamefont {Cleland},\ and\ \citenamefont
  {Bachtold}}]{prx-pistolesi-2021}%
  \BibitemOpen
  \bibfield  {author} {\bibinfo {author} {\bibfnamefont {F.}~\bibnamefont
  {Pistolesi}}, \bibinfo {author} {\bibfnamefont {A.~N.}\ \bibnamefont
  {Cleland}},\ and\ \bibinfo {author} {\bibfnamefont {A.}~\bibnamefont
  {Bachtold}},\ }\bibfield  {title} {\bibinfo {title} {Proposal for a
  nanomechanical qubit},\ }\href {https://doi.org/10.1103/PhysRevX.11.031027}
  {\bibfield  {journal} {\bibinfo  {journal} {Physical Review X}\ }\textbf
  {\bibinfo {volume} {11}},\ \bibinfo {pages} {031027} (\bibinfo {year}
  {2021})}\BibitemShut {NoStop}%
\bibitem [{\citenamefont {Wen}\ \emph {et~al.}(2020)\citenamefont {Wen},
  \citenamefont {Ares}, \citenamefont {Schupp}, \citenamefont {Pei},
  \citenamefont {Briggs},\ and\ \citenamefont {Laird}}]{nphys-wen-2020}%
  \BibitemOpen
  \bibfield  {author} {\bibinfo {author} {\bibfnamefont {Y.}~\bibnamefont
  {Wen}}, \bibinfo {author} {\bibfnamefont {N.}~\bibnamefont {Ares}}, \bibinfo
  {author} {\bibfnamefont {F.~J.}\ \bibnamefont {Schupp}}, \bibinfo {author}
  {\bibfnamefont {T.}~\bibnamefont {Pei}}, \bibinfo {author} {\bibfnamefont
  {G.~a.~D.}\ \bibnamefont {Briggs}},\ and\ \bibinfo {author} {\bibfnamefont
  {E.~A.}\ \bibnamefont {Laird}},\ }\bibfield  {title} {\bibinfo {title} {A
  coherent nanomechanical oscillator driven by single-electron tunnelling},\
  }\href {https://doi.org/10.1038/s41567-019-0683-5} {\bibfield  {journal}
  {\bibinfo  {journal} {Nature Physics}\ }\textbf {\bibinfo {volume} {16}},\
  \bibinfo {pages} {75} (\bibinfo {year} {2020})}\BibitemShut {NoStop}%
\bibitem [{\citenamefont {Vigneau}\ \emph {et~al.}(2022)\citenamefont
  {Vigneau}, \citenamefont {Monsel}, \citenamefont {Tabanera}, \citenamefont
  {Aggarwal}, \citenamefont {Bresque}, \citenamefont {Fedele}, \citenamefont
  {Cerisola}, \citenamefont {Briggs}, \citenamefont {Anders}, \citenamefont
  {Parrondo}, \citenamefont {Auff{\`e}ves},\ and\ \citenamefont
  {Ares}}]{prr-vigneau-2022}%
  \BibitemOpen
  \bibfield  {author} {\bibinfo {author} {\bibfnamefont {F.}~\bibnamefont
  {Vigneau}}, \bibinfo {author} {\bibfnamefont {J.}~\bibnamefont {Monsel}},
  \bibinfo {author} {\bibfnamefont {J.}~\bibnamefont {Tabanera}}, \bibinfo
  {author} {\bibfnamefont {K.}~\bibnamefont {Aggarwal}}, \bibinfo {author}
  {\bibfnamefont {L.}~\bibnamefont {Bresque}}, \bibinfo {author} {\bibfnamefont
  {F.}~\bibnamefont {Fedele}}, \bibinfo {author} {\bibfnamefont
  {F.}~\bibnamefont {Cerisola}}, \bibinfo {author} {\bibfnamefont {G.~A.~D.}\
  \bibnamefont {Briggs}}, \bibinfo {author} {\bibfnamefont {J.}~\bibnamefont
  {Anders}}, \bibinfo {author} {\bibfnamefont {J.~M.~R.}\ \bibnamefont
  {Parrondo}}, \bibinfo {author} {\bibfnamefont {A.}~\bibnamefont
  {Auff{\`e}ves}},\ and\ \bibinfo {author} {\bibfnamefont {N.}~\bibnamefont
  {Ares}},\ }\bibfield  {title} {\bibinfo {title} {Ultrastrong coupling between
  electron tunneling and mechanical motion},\ }\href
  {https://doi.org/10.1103/PhysRevResearch.4.043168} {\bibfield  {journal}
  {\bibinfo  {journal} {Physical Review Research}\ }\textbf {\bibinfo {volume}
  {4}},\ \bibinfo {pages} {043168} (\bibinfo {year} {2022})}\BibitemShut
  {NoStop}%
\bibitem [{\citenamefont {Postma}\ \emph {et~al.}(2005)\citenamefont {Postma},
  \citenamefont {Kozinsky}, \citenamefont {Husain},\ and\ \citenamefont
  {Roukes}}]{apl-postma-2005}%
  \BibitemOpen
  \bibfield  {author} {\bibinfo {author} {\bibfnamefont {H.~W.~{\relax Ch}.}\
  \bibnamefont {Postma}}, \bibinfo {author} {\bibfnamefont {I.}~\bibnamefont
  {Kozinsky}}, \bibinfo {author} {\bibfnamefont {A.}~\bibnamefont {Husain}},\
  and\ \bibinfo {author} {\bibfnamefont {M.~L.}\ \bibnamefont {Roukes}},\
  }\bibfield  {title} {\bibinfo {title} {Dynamic range of nanotube- and
  nanowire-based electromechanical systems},\ }\href
  {https://doi.org/10.1063/1.1929098} {\bibfield  {journal} {\bibinfo
  {journal} {Applied Physics Letters}\ }\textbf {\bibinfo {volume} {86}},\
  \bibinfo {pages} {223105} (\bibinfo {year} {2005})}\BibitemShut {NoStop}%
\bibitem [{\citenamefont {Duffing}(1918)}]{book-duffing-1918}%
  \BibitemOpen
  \bibfield  {author} {\bibinfo {author} {\bibfnamefont {G.}~\bibnamefont
  {Duffing}},\ }\href@noop {} {\emph {\bibinfo {title} {Erzwungene
  {{Schwingungen}} bei ver{\"a}nderlicher {{Eigenfrequenz}} und ihre technische
  {{Bedeutung}}}}}\ (\bibinfo  {publisher} {Friedrich Vieweg \& Sohn},\
  \bibinfo {address} {Braunschweig},\ \bibinfo {year} {1918})\BibitemShut
  {NoStop}%
\bibitem [{\citenamefont {Blien}\ \emph {et~al.}(2018)\citenamefont {Blien},
  \citenamefont {Steger}, \citenamefont {Albang}, \citenamefont {Paradiso},\
  and\ \citenamefont {H{\"u}ttel}}]{forktransfer}%
  \BibitemOpen
  \bibfield  {author} {\bibinfo {author} {\bibfnamefont {S.}~\bibnamefont
  {Blien}}, \bibinfo {author} {\bibfnamefont {P.}~\bibnamefont {Steger}},
  \bibinfo {author} {\bibfnamefont {A.}~\bibnamefont {Albang}}, \bibinfo
  {author} {\bibfnamefont {N.}~\bibnamefont {Paradiso}},\ and\ \bibinfo
  {author} {\bibfnamefont {A.~K.}\ \bibnamefont {H{\"u}ttel}},\ }\bibfield
  {title} {\bibinfo {title} {Quartz tuning-fork based carbon nanotube transfer
  into quantum device geometries},\ }\href
  {https://doi.org/10.1002/pssb.201800118} {\bibfield  {journal} {\bibinfo
  {journal} {Physica Status Solidi B}\ }\textbf {\bibinfo {volume} {255}},\
  \bibinfo {pages} {1800118} (\bibinfo {year} {2018})}\BibitemShut {NoStop}%
\bibitem [{\citenamefont {Kellner}\ \emph {et~al.}(2023)\citenamefont
  {Kellner}, \citenamefont {H{\"u}ttner}, \citenamefont {Will}, \citenamefont
  {Hakonen},\ and\ \citenamefont {H{\"u}ttel}}]{stepwisefab}%
  \BibitemOpen
  \bibfield  {author} {\bibinfo {author} {\bibfnamefont {N.}~\bibnamefont
  {Kellner}}, \bibinfo {author} {\bibfnamefont {N.}~\bibnamefont
  {H{\"u}ttner}}, \bibinfo {author} {\bibfnamefont {M.}~\bibnamefont {Will}},
  \bibinfo {author} {\bibfnamefont {P.}~\bibnamefont {Hakonen}},\ and\ \bibinfo
  {author} {\bibfnamefont {A.~K.}\ \bibnamefont {H{\"u}ttel}},\ }\bibfield
  {title} {\bibinfo {title} {Stepwise fabrication and optimization of coplanar
  waveguide resonator hybrid devices},\ }\href
  {https://doi.org/10.1002/pssb.202300187} {\bibfield  {journal} {\bibinfo
  {journal} {Physica Status Solidi B}\ }\textbf {\bibinfo {volume} {260}},\
  \bibinfo {pages} {2300187} (\bibinfo {year} {2023})}\BibitemShut {NoStop}%
\bibitem [{\citenamefont {H{\"u}ttner}\ \emph {et~al.}(2023)\citenamefont
  {H{\"u}ttner}, \citenamefont {Blien}, \citenamefont {Steger}, \citenamefont
  {Loh}, \citenamefont {Graaf},\ and\ \citenamefont
  {H{\"u}ttel}}]{modelingomit}%
  \BibitemOpen
  \bibfield  {author} {\bibinfo {author} {\bibfnamefont {N.}~\bibnamefont
  {H{\"u}ttner}}, \bibinfo {author} {\bibfnamefont {S.}~\bibnamefont {Blien}},
  \bibinfo {author} {\bibfnamefont {P.}~\bibnamefont {Steger}}, \bibinfo
  {author} {\bibfnamefont {A.}~\bibnamefont {Loh}}, \bibinfo {author}
  {\bibfnamefont {R.}~\bibnamefont {Graaf}},\ and\ \bibinfo {author}
  {\bibfnamefont {A.}~\bibnamefont {H{\"u}ttel}},\ }\bibfield  {title}
  {\bibinfo {title} {Optomechanical coupling and damping of a carbon nanotube
  quantum dot},\ }\href {https://doi.org/10.1103/PhysRevApplied.20.064019}
  {\bibfield  {journal} {\bibinfo  {journal} {Physical Review Applied}\
  }\textbf {\bibinfo {volume} {20}},\ \bibinfo {pages} {064019} (\bibinfo
  {year} {2023})}\BibitemShut {NoStop}%
\bibitem [{\citenamefont {G{\"o}tz}\ \emph {et~al.}(2018)\citenamefont
  {G{\"o}tz}, \citenamefont {Schmid}, \citenamefont {Schupp}, \citenamefont
  {Stiller}, \citenamefont {Strunk},\ and\ \citenamefont
  {H{\"u}ttel}}]{kondocharge}%
  \BibitemOpen
  \bibfield  {author} {\bibinfo {author} {\bibfnamefont {K.~J.~G.}\
  \bibnamefont {G{\"o}tz}}, \bibinfo {author} {\bibfnamefont {D.~R.}\
  \bibnamefont {Schmid}}, \bibinfo {author} {\bibfnamefont {F.~J.}\
  \bibnamefont {Schupp}}, \bibinfo {author} {\bibfnamefont {P.~L.}\
  \bibnamefont {Stiller}}, \bibinfo {author} {\bibfnamefont {{\relax
  Ch}.}~\bibnamefont {Strunk}},\ and\ \bibinfo {author} {\bibfnamefont {A.~K.}\
  \bibnamefont {H{\"u}ttel}},\ }\bibfield  {title} {\bibinfo {title}
  {Nanomechanical characterization of the {{Kondo}} charge dynamics in a carbon
  nanotube},\ }\href {https://doi.org/10.1103/PhysRevLett.120.246802}
  {\bibfield  {journal} {\bibinfo  {journal} {Physical Review Letters}\
  }\textbf {\bibinfo {volume} {120}},\ \bibinfo {pages} {246802} (\bibinfo
  {year} {2018})}\BibitemShut {NoStop}%
\bibitem [{\citenamefont {Chen}\ \emph {et~al.}(2009)\citenamefont {Chen},
  \citenamefont {Rosenblatt}, \citenamefont {Bolotin}, \citenamefont {Kalb},
  \citenamefont {Kim}, \citenamefont {Kymissis}, \citenamefont {Stormer},
  \citenamefont {Heinz},\ and\ \citenamefont {Hone}}]{nn-chen-2009}%
  \BibitemOpen
  \bibfield  {author} {\bibinfo {author} {\bibfnamefont {C.}~\bibnamefont
  {Chen}}, \bibinfo {author} {\bibfnamefont {S.}~\bibnamefont {Rosenblatt}},
  \bibinfo {author} {\bibfnamefont {K.~I.}\ \bibnamefont {Bolotin}}, \bibinfo
  {author} {\bibfnamefont {W.}~\bibnamefont {Kalb}}, \bibinfo {author}
  {\bibfnamefont {P.}~\bibnamefont {Kim}}, \bibinfo {author} {\bibfnamefont
  {I.}~\bibnamefont {Kymissis}}, \bibinfo {author} {\bibfnamefont {H.~L.}\
  \bibnamefont {Stormer}}, \bibinfo {author} {\bibfnamefont {T.~F.}\
  \bibnamefont {Heinz}},\ and\ \bibinfo {author} {\bibfnamefont
  {J.}~\bibnamefont {Hone}},\ }\bibfield  {title} {\bibinfo {title}
  {Performance of monolayer graphene nanomechanical resonators with electrical
  readout},\ }\href {https://doi.org/10.1038/nnano.2009.267} {\bibfield
  {journal} {\bibinfo  {journal} {Nature Nanotechnology}\ }\textbf {\bibinfo
  {volume} {4}},\ \bibinfo {pages} {861} (\bibinfo {year} {2009})}\BibitemShut
  {NoStop}%
\bibitem [{\citenamefont {Castro~Neto}\ \emph {et~al.}(2009)\citenamefont
  {Castro~Neto}, \citenamefont {Guinea}, \citenamefont {Peres}, \citenamefont
  {Novoselov},\ and\ \citenamefont {Geim}}]{rmp-castroneto-2009}%
  \BibitemOpen
  \bibfield  {author} {\bibinfo {author} {\bibfnamefont {A.~H.}\ \bibnamefont
  {Castro~Neto}}, \bibinfo {author} {\bibfnamefont {F.}~\bibnamefont {Guinea}},
  \bibinfo {author} {\bibfnamefont {N.~M.~R.}\ \bibnamefont {Peres}}, \bibinfo
  {author} {\bibfnamefont {K.~S.}\ \bibnamefont {Novoselov}},\ and\ \bibinfo
  {author} {\bibfnamefont {A.~K.}\ \bibnamefont {Geim}},\ }\bibfield  {title}
  {\bibinfo {title} {The electronic properties of graphene},\ }\href
  {https://doi.org/10.1103/RevModPhys.81.109} {\bibfield  {journal} {\bibinfo
  {journal} {Reviews of Modern Physics}\ }\textbf {\bibinfo {volume} {81}},\
  \bibinfo {pages} {109} (\bibinfo {year} {2009})}\BibitemShut {NoStop}%
\bibitem [{\citenamefont {Lifshitz}\ and\ \citenamefont
  {Cross}(2008)}]{rondac-lifshitz-2008}%
  \BibitemOpen
  \bibfield  {author} {\bibinfo {author} {\bibfnamefont {R.}~\bibnamefont
  {Lifshitz}}\ and\ \bibinfo {author} {\bibfnamefont {M.~C.}\ \bibnamefont
  {Cross}},\ }\bibinfo {title} {Nonlinear dynamics of nanomechanical and
  micromechanical resonators},\ in\ \href
  {https://doi.org/10.1002/9783527626359.ch1} {\emph {\bibinfo {booktitle}
  {Reviews of Nonlinear Dynamics and Complexity}}}\ (\bibinfo  {publisher}
  {John Wiley \& Sons, Ltd},\ \bibinfo {year} {2008})\ Chap.~\bibinfo {chapter}
  {1}, pp.\ \bibinfo {pages} {1--52}\BibitemShut {NoStop}%
\bibitem [{\citenamefont {Micchi}\ \emph {et~al.}(2015)\citenamefont {Micchi},
  \citenamefont {Avriller},\ and\ \citenamefont {Pistolesi}}]{prl-micchi-2015}%
  \BibitemOpen
  \bibfield  {author} {\bibinfo {author} {\bibfnamefont {G.}~\bibnamefont
  {Micchi}}, \bibinfo {author} {\bibfnamefont {R.}~\bibnamefont {Avriller}},\
  and\ \bibinfo {author} {\bibfnamefont {F.}~\bibnamefont {Pistolesi}},\
  }\bibfield  {title} {\bibinfo {title} {Mechanical {{Signatures}} of the
  {{Current Blockade Instability}} in {{Suspended Carbon Nanotubes}}},\ }\href
  {https://doi.org/10.1103/PhysRevLett.115.206802} {\bibfield  {journal}
  {\bibinfo  {journal} {Physical Review Letters}\ }\textbf {\bibinfo {volume}
  {115}},\ \bibinfo {pages} {206802} (\bibinfo {year} {2015})}\BibitemShut
  {NoStop}%
\bibitem [{\citenamefont {Micchi}\ \emph {et~al.}(2016)\citenamefont {Micchi},
  \citenamefont {Avriller},\ and\ \citenamefont {Pistolesi}}]{prb-micchi-2016}%
  \BibitemOpen
  \bibfield  {author} {\bibinfo {author} {\bibfnamefont {G.}~\bibnamefont
  {Micchi}}, \bibinfo {author} {\bibfnamefont {R.}~\bibnamefont {Avriller}},\
  and\ \bibinfo {author} {\bibfnamefont {F.}~\bibnamefont {Pistolesi}},\
  }\bibfield  {title} {\bibinfo {title} {Electromechanical transition in
  quantum dots},\ }\href {https://doi.org/10.1103/PhysRevB.94.125417}
  {\bibfield  {journal} {\bibinfo  {journal} {Physical Review B}\ }\textbf
  {\bibinfo {volume} {94}},\ \bibinfo {pages} {125417} (\bibinfo {year}
  {2016})}\BibitemShut {NoStop}%
\bibitem [{\citenamefont {Jones}\ and\ \citenamefont
  {Trefan}(2001)}]{ajp-jones-2001}%
  \BibitemOpen
  \bibfield  {author} {\bibinfo {author} {\bibfnamefont {B.~K.}\ \bibnamefont
  {Jones}}\ and\ \bibinfo {author} {\bibfnamefont {G.}~\bibnamefont {Trefan}},\
  }\bibfield  {title} {\bibinfo {title} {The {{Duffing}} oscillator: {{A}}
  precise electronic analog chaos demonstrator for the undergraduate
  laboratory},\ }\href {https://doi.org/10.1119/1.1336838} {\bibfield
  {journal} {\bibinfo  {journal} {American Journal of Physics}\ }\textbf
  {\bibinfo {volume} {69}},\ \bibinfo {pages} {464} (\bibinfo {year}
  {2001})}\BibitemShut {NoStop}%
\bibitem [{\citenamefont {Brennan}\ \emph {et~al.}(2008)\citenamefont
  {Brennan}, \citenamefont {Kovacic}, \citenamefont {Carrella},\ and\
  \citenamefont {Waters}}]{josav-brennan-2008}%
  \BibitemOpen
  \bibfield  {author} {\bibinfo {author} {\bibfnamefont {M.}~\bibnamefont
  {Brennan}}, \bibinfo {author} {\bibfnamefont {I.}~\bibnamefont {Kovacic}},
  \bibinfo {author} {\bibfnamefont {A.}~\bibnamefont {Carrella}},\ and\
  \bibinfo {author} {\bibfnamefont {T.}~\bibnamefont {Waters}},\ }\bibfield
  {title} {\bibinfo {title} {On the jump-up and jump-down frequencies of the
  {{Duffing}} oscillator},\ }\href {https://doi.org/10.1016/j.jsv.2008.04.032}
  {\bibfield  {journal} {\bibinfo  {journal} {Journal of Sound and Vibration}\
  }\textbf {\bibinfo {volume} {318}},\ \bibinfo {pages} {1250} (\bibinfo {year}
  {2008})}\BibitemShut {NoStop}%
\bibitem [{\citenamefont {Warminski}\ and\ \citenamefont
  {Kecik}(2009)}]{josav-warminski-2009}%
  \BibitemOpen
  \bibfield  {author} {\bibinfo {author} {\bibfnamefont {J.}~\bibnamefont
  {Warminski}}\ and\ \bibinfo {author} {\bibfnamefont {K.}~\bibnamefont
  {Kecik}},\ }\bibfield  {title} {\bibinfo {title} {Instabilities in the main
  parametric resonance area of a mechanical system with a pendulum},\ }\href
  {https://doi.org/10.1016/j.jsv.2008.06.042} {\bibfield  {journal} {\bibinfo
  {journal} {Journal of Sound and Vibration}\ }\bibinfo {series} {Special Issue
  from the {{Second International Conference}} on {{Nonlinear Dynamics}},
  {{Kharkov}}, {{Ukraine}}, 25-28 {{September}} 2007, Held in Honour of the
  150th Anniversary of {{Alexander Mikhailovich Lyapunov}}},\ \textbf {\bibinfo
  {volume} {322}},\ \bibinfo {pages} {612} (\bibinfo {year}
  {2009})}\BibitemShut {NoStop}%
\bibitem [{\citenamefont {Clerc}\ \emph {et~al.}(2018)\citenamefont {Clerc},
  \citenamefont {Coulibaly}, \citenamefont {Ferr{\'e}},\ and\ \citenamefont
  {Rojas}}]{c-clerc-2018}%
  \BibitemOpen
  \bibfield  {author} {\bibinfo {author} {\bibfnamefont {M.~G.}\ \bibnamefont
  {Clerc}}, \bibinfo {author} {\bibfnamefont {S.}~\bibnamefont {Coulibaly}},
  \bibinfo {author} {\bibfnamefont {M.~A.}\ \bibnamefont {Ferr{\'e}}},\ and\
  \bibinfo {author} {\bibfnamefont {R.~G.}\ \bibnamefont {Rojas}},\ }\bibfield
  {title} {\bibinfo {title} {Chimera states in a {{Duffing}} oscillators chain
  coupled to nearest neighbors},\ }\href {https://doi.org/10.1063/1.5025038}
  {\bibfield  {journal} {\bibinfo  {journal} {Chaos (Woodbury, N.Y.)}\ }\textbf
  {\bibinfo {volume} {28}},\ \bibinfo {pages} {083126} (\bibinfo {year}
  {2018})}\BibitemShut {NoStop}%
\bibitem [{\citenamefont {Ramlan}\ \emph {et~al.}(2016)\citenamefont {Ramlan},
  \citenamefont {Brennan}, \citenamefont {Kovacic}, \citenamefont {Mace},\ and\
  \citenamefont {Burrow}}]{cinsans-ramlan-2016}%
  \BibitemOpen
  \bibfield  {author} {\bibinfo {author} {\bibfnamefont {R.}~\bibnamefont
  {Ramlan}}, \bibinfo {author} {\bibfnamefont {M.~J.}\ \bibnamefont {Brennan}},
  \bibinfo {author} {\bibfnamefont {I.}~\bibnamefont {Kovacic}}, \bibinfo
  {author} {\bibfnamefont {B.~R.}\ \bibnamefont {Mace}},\ and\ \bibinfo
  {author} {\bibfnamefont {S.~G.}\ \bibnamefont {Burrow}},\ }\bibfield  {title}
  {\bibinfo {title} {Exploiting knowledge of jump-up and jump-down frequencies
  to determine the parameters of a {{Duffing}} oscillator},\ }\href
  {https://doi.org/10.1016/j.cnsns.2016.01.017} {\bibfield  {journal} {\bibinfo
   {journal} {Communications in Nonlinear Science and Numerical Simulation}\
  }\textbf {\bibinfo {volume} {37}},\ \bibinfo {pages} {282} (\bibinfo {year}
  {2016})}\BibitemShut {NoStop}%
\bibitem [{\citenamefont {Ghouli}\ \emph {et~al.}(2017)\citenamefont {Ghouli},
  \citenamefont {Hamdi}, \citenamefont {Lakrad},\ and\ \citenamefont
  {Belhaq}}]{josav-ghouli-2017}%
  \BibitemOpen
  \bibfield  {author} {\bibinfo {author} {\bibfnamefont {Z.}~\bibnamefont
  {Ghouli}}, \bibinfo {author} {\bibfnamefont {M.}~\bibnamefont {Hamdi}},
  \bibinfo {author} {\bibfnamefont {F.}~\bibnamefont {Lakrad}},\ and\ \bibinfo
  {author} {\bibfnamefont {M.}~\bibnamefont {Belhaq}},\ }\bibfield  {title}
  {\bibinfo {title} {Quasiperiodic energy harvesting in a forced and delayed
  {{Duffing}} harvester device},\ }\href
  {https://doi.org/10.1016/j.jsv.2017.07.005} {\bibfield  {journal} {\bibinfo
  {journal} {Journal of Sound and Vibration}\ }\textbf {\bibinfo {volume}
  {407}},\ \bibinfo {pages} {271} (\bibinfo {year} {2017})}\BibitemShut
  {NoStop}%
\bibitem [{\citenamefont {Catalini}\ \emph {et~al.}(2021)\citenamefont
  {Catalini}, \citenamefont {Rossi}, \citenamefont {Langman},\ and\
  \citenamefont {Schliesser}}]{prl-catalini-2021}%
  \BibitemOpen
  \bibfield  {author} {\bibinfo {author} {\bibfnamefont {L.}~\bibnamefont
  {Catalini}}, \bibinfo {author} {\bibfnamefont {M.}~\bibnamefont {Rossi}},
  \bibinfo {author} {\bibfnamefont {E.~C.}\ \bibnamefont {Langman}},\ and\
  \bibinfo {author} {\bibfnamefont {A.}~\bibnamefont {Schliesser}},\ }\bibfield
   {title} {\bibinfo {title} {Modeling and {{Observation}} of {{Nonlinear
  Damping}} in {{Dissipation-Diluted Nanomechanical Resonators}}},\ }\href
  {https://doi.org/10.1103/PhysRevLett.126.174101} {\bibfield  {journal}
  {\bibinfo  {journal} {Physical Review Letters}\ }\textbf {\bibinfo {volume}
  {126}},\ \bibinfo {pages} {174101} (\bibinfo {year} {2021})}\BibitemShut
  {NoStop}%
\bibitem [{\citenamefont {Kaisar}\ \emph {et~al.}(2022)\citenamefont {Kaisar},
  \citenamefont {Lee}, \citenamefont {Li}, \citenamefont {Shaw},\ and\
  \citenamefont {Feng}}]{nl-kaisar-2022}%
  \BibitemOpen
  \bibfield  {author} {\bibinfo {author} {\bibfnamefont {T.}~\bibnamefont
  {Kaisar}}, \bibinfo {author} {\bibfnamefont {J.}~\bibnamefont {Lee}},
  \bibinfo {author} {\bibfnamefont {D.}~\bibnamefont {Li}}, \bibinfo {author}
  {\bibfnamefont {S.~W.}\ \bibnamefont {Shaw}},\ and\ \bibinfo {author}
  {\bibfnamefont {P.~X.-L.}\ \bibnamefont {Feng}},\ }\bibfield  {title}
  {\bibinfo {title} {Nonlinear {{Stiffness}} and {{Nonlinear Damping}} in
  {{Atomically Thin MoS}}{\textsubscript{2}} {{Nanomechanical Resonators}}},\
  }\href {https://doi.org/10.1021/acs.nanolett.2c02629} {\bibfield  {journal}
  {\bibinfo  {journal} {Nano Letters}\ }\textbf {\bibinfo {volume} {22}},\
  \bibinfo {pages} {9831} (\bibinfo {year} {2022})}\BibitemShut {NoStop}%
\bibitem [{\citenamefont {Wawrzynski}(2022)}]{sr-wawrzynski-2022}%
  \BibitemOpen
  \bibfield  {author} {\bibinfo {author} {\bibfnamefont {W.}~\bibnamefont
  {Wawrzynski}},\ }\bibfield  {title} {\bibinfo {title} {The origin point of
  the unstable solution area of a forced softening {{Duffing}} oscillator},\
  }\href {https://doi.org/10.1038/s41598-022-07932-8} {\bibfield  {journal}
  {\bibinfo  {journal} {Scientific Reports}\ }\textbf {\bibinfo {volume}
  {12}},\ \bibinfo {pages} {4518} (\bibinfo {year} {2022})}\BibitemShut
  {NoStop}%
\bibitem [{\citenamefont {Wang}\ \emph {et~al.}(2021)\citenamefont {Wang},
  \citenamefont {Cong}, \citenamefont {Zhu}, \citenamefont {Yuan},
  \citenamefont {Lin}, \citenamefont {Zhao}, \citenamefont {Bai}, \citenamefont
  {Liang}, \citenamefont {Sun}, \citenamefont {Deng},\ and\ \citenamefont
  {Jiang}}]{wang_visualizing_2021}%
  \BibitemOpen
  \bibfield  {author} {\bibinfo {author} {\bibfnamefont {X.}~\bibnamefont
  {Wang}}, \bibinfo {author} {\bibfnamefont {L.}~\bibnamefont {Cong}}, \bibinfo
  {author} {\bibfnamefont {D.}~\bibnamefont {Zhu}}, \bibinfo {author}
  {\bibfnamefont {Z.}~\bibnamefont {Yuan}}, \bibinfo {author} {\bibfnamefont
  {X.}~\bibnamefont {Lin}}, \bibinfo {author} {\bibfnamefont {W.}~\bibnamefont
  {Zhao}}, \bibinfo {author} {\bibfnamefont {Z.}~\bibnamefont {Bai}}, \bibinfo
  {author} {\bibfnamefont {W.}~\bibnamefont {Liang}}, \bibinfo {author}
  {\bibfnamefont {X.}~\bibnamefont {Sun}}, \bibinfo {author} {\bibfnamefont
  {G.-W.}\ \bibnamefont {Deng}},\ and\ \bibinfo {author} {\bibfnamefont
  {K.}~\bibnamefont {Jiang}},\ }\bibfield  {title} {\bibinfo {title}
  {Visualizing nonlinear resonance in nanomechanical systems via
  single-electron tunneling},\ }\href
  {https://doi.org/10.1007/s12274-020-3165-2} {\bibfield  {journal} {\bibinfo
  {journal} {Nano Research}\ }\textbf {\bibinfo {volume} {14}},\ \bibinfo
  {pages} {1156} (\bibinfo {year} {2021})}\BibitemShut {NoStop}%
\bibitem [{\citenamefont {Jordan}\ and\ \citenamefont
  {Smith}(2023)}]{-jordan-2023}%
  \BibitemOpen
  \bibfield  {author} {\bibinfo {author} {\bibfnamefont {D.~W.}\ \bibnamefont
  {Jordan}}\ and\ \bibinfo {author} {\bibfnamefont {P.}~\bibnamefont {Smith}},\
  }\href {https://doi.org/10.1093/oso/9780199208241.001.0001} {\emph {\bibinfo
  {title} {Nonlinear Ordinary Differential Equations: An Introduction for
  Scientists and Engineers}}},\ \bibinfo {edition} {4th}\ ed.,\ Oxford
  Scholarship Online\ (\bibinfo  {publisher} {Oxford University Press},\
  \bibinfo {address} {Oxford},\ \bibinfo {year} {2023})\BibitemShut {NoStop}%
\bibitem [{\citenamefont {Willick}\ \emph {et~al.}(2017)\citenamefont
  {Willick}, \citenamefont {Tang},\ and\ \citenamefont
  {Baugh}}]{willick_probing_2017}%
  \BibitemOpen
  \bibfield  {author} {\bibinfo {author} {\bibfnamefont {K.}~\bibnamefont
  {Willick}}, \bibinfo {author} {\bibfnamefont {X.~S.}\ \bibnamefont {Tang}},\
  and\ \bibinfo {author} {\bibfnamefont {J.}~\bibnamefont {Baugh}},\ }\bibfield
   {title} {\bibinfo {title} {Probing the non-linear transient response of a
  carbon nanotube mechanical oscillator},\ }\href
  {https://doi.org/10.1063/1.4991412} {\bibfield  {journal} {\bibinfo
  {journal} {Applied Physics Letters}\ }\textbf {\bibinfo {volume} {111}},\
  \bibinfo {pages} {223108} (\bibinfo {year} {2017})}\BibitemShut {NoStop}%
\bibitem [{\citenamefont {Luo}\ \emph {et~al.}(2017)\citenamefont {Luo},
  \citenamefont {Zhang}, \citenamefont {Deng}, \citenamefont {Li},
  \citenamefont {Cao}, \citenamefont {Xiao}, \citenamefont {Guo},\ and\
  \citenamefont {Guo}}]{luo_coupling_2017}%
  \BibitemOpen
  \bibfield  {author} {\bibinfo {author} {\bibfnamefont {G.}~\bibnamefont
  {Luo}}, \bibinfo {author} {\bibfnamefont {Z.-Z.}\ \bibnamefont {Zhang}},
  \bibinfo {author} {\bibfnamefont {G.-W.}\ \bibnamefont {Deng}}, \bibinfo
  {author} {\bibfnamefont {H.-O.}\ \bibnamefont {Li}}, \bibinfo {author}
  {\bibfnamefont {G.}~\bibnamefont {Cao}}, \bibinfo {author} {\bibfnamefont
  {M.}~\bibnamefont {Xiao}}, \bibinfo {author} {\bibfnamefont {G.-C.}\
  \bibnamefont {Guo}},\ and\ \bibinfo {author} {\bibfnamefont {G.-P.}\
  \bibnamefont {Guo}},\ }\bibfield  {title} {\bibinfo {title} {Coupling
  graphene nanomechanical motion to a single-electron transistor},\ }\href
  {https://doi.org/10.1039/c6nr09768e} {\bibfield  {journal} {\bibinfo
  {journal} {Nanoscale}\ }\textbf {\bibinfo {volume} {9}},\ \bibinfo {pages}
  {5608} (\bibinfo {year} {2017})}\BibitemShut {NoStop}%
\bibitem [{\citenamefont {Eichler}\ \emph {et~al.}(2011)\citenamefont
  {Eichler}, \citenamefont {Moser}, \citenamefont {Chaste}, \citenamefont
  {Zdrojek}, \citenamefont {{Wilson-Rae}},\ and\ \citenamefont
  {Bachtold}}]{nnano-eichler-2011}%
  \BibitemOpen
  \bibfield  {author} {\bibinfo {author} {\bibfnamefont {A.}~\bibnamefont
  {Eichler}}, \bibinfo {author} {\bibfnamefont {J.}~\bibnamefont {Moser}},
  \bibinfo {author} {\bibfnamefont {J.}~\bibnamefont {Chaste}}, \bibinfo
  {author} {\bibfnamefont {M.}~\bibnamefont {Zdrojek}}, \bibinfo {author}
  {\bibfnamefont {I.}~\bibnamefont {{Wilson-Rae}}},\ and\ \bibinfo {author}
  {\bibfnamefont {A.}~\bibnamefont {Bachtold}},\ }\bibfield  {title} {\bibinfo
  {title} {Nonlinear damping in mechanical resonators made from carbon
  nanotubes and graphene},\ }\href {https://doi.org/10.1038/nnano.2011.71}
  {\bibfield  {journal} {\bibinfo  {journal} {Nature Nanotechnology}\ }\textbf
  {\bibinfo {volume} {6}},\ \bibinfo {pages} {339} (\bibinfo {year}
  {2011})}\BibitemShut {NoStop}%
\bibitem [{\citenamefont {Reinhardt}\ \emph {et~al.}(2019)\citenamefont
  {Reinhardt}, \citenamefont {Butschkow}, \citenamefont {Geissler},
  \citenamefont {Dirnaichner}, \citenamefont {Olbrich}, \citenamefont {Lane},
  \citenamefont {Schr{\"o}er},\ and\ \citenamefont
  {H{\"u}ttel}}]{labmeasurement}%
  \BibitemOpen
  \bibfield  {author} {\bibinfo {author} {\bibfnamefont {S.}~\bibnamefont
  {Reinhardt}}, \bibinfo {author} {\bibfnamefont {C.}~\bibnamefont
  {Butschkow}}, \bibinfo {author} {\bibfnamefont {S.}~\bibnamefont {Geissler}},
  \bibinfo {author} {\bibfnamefont {A.}~\bibnamefont {Dirnaichner}}, \bibinfo
  {author} {\bibfnamefont {F.}~\bibnamefont {Olbrich}}, \bibinfo {author}
  {\bibfnamefont {C.}~\bibnamefont {Lane}}, \bibinfo {author} {\bibfnamefont
  {D.}~\bibnamefont {Schr{\"o}er}},\ and\ \bibinfo {author} {\bibfnamefont
  {A.~K.}\ \bibnamefont {H{\"u}ttel}},\ }\bibfield  {title} {\bibinfo {title}
  {Lab::{{Measurement}} --- a portable and extensible framework for controlling
  lab equipment and conducting measurements},\ }\href
  {https://doi.org/10.1016/j.cpc.2018.07.024} {\bibfield  {journal} {\bibinfo
  {journal} {Computer Physics Communications}\ }\textbf {\bibinfo {volume}
  {234}},\ \bibinfo {pages} {216} (\bibinfo {year} {2019})}\BibitemShut
  {NoStop}%
\end{thebibliography}%

\end{document}